\documentstyle[aps,prl,preprint,floats,epsfig]{revtex}
\begin{document}
\newcommand{\mev   }{\mbox{\rm MeV}}
\newcommand{\mevc  }{\mbox{\rm MeV/$c$}}
\newcommand{\mevcsq}{\mbox{\rm MeV/$c^2$}}
\newcommand{\gev   }{\mbox{\rm GeV}}
\newcommand{\gevc  }{\mbox{\rm GeV/$c$}}
\newcommand{\gevcsq}{\mbox{\rm GeV/$c^2$}}
\newcommand{\decays}{\mbox{$\rightarrow$}}
\newcommand{\xp}{\mbox{$x_p$}} 
\newcommand{\q}{\mbox{$q$}}
\newcommand{\uq}{\mbox{$u$}}
\newcommand{\dq}{\mbox{$d$}}
\newcommand{\sq}{\mbox{$s$}}
\newcommand{\cq}{\mbox{$c$}}
\newcommand{\cbar}{\mbox{$\overline{c}$}}
\newcommand{\bq}{\mbox{$b$}}
\newcommand{\tq}{\mbox{$t$}}
\newcommand{\phot}{\mbox{$\gamma$}}
\newcommand{\kmi}{\mbox{$K^-$}}
\newcommand{\kpl}{\mbox{$K^+$}}
\newcommand{\lz}{\mbox{$\Lambda$}}
\newcommand{\sigpl}{\mbox{$\Sigma^+$}}
\newcommand{\lcpl}{\mbox{$\Lambda^+_c$}}
\newcommand{\casmi}{\mbox{$\Xi^-$}}
\newcommand{\casz}{\mbox{$\Xi^0$}}
\newcommand{\omgmi}{\mbox{$\Omega^-$}}
\newcommand{\casc}{\mbox{$\Xi_c$}}
\newcommand{\cascz}{\mbox{$\Xi^0_c$}}
\newcommand{\cascpl}{\mbox{$\Xi^+_c$}}
\newcommand{\omegac}{\mbox{$\Omega_c^0$}}
\newcommand{\mcol}[3]{\multicolumn{#1}{#2}{#3}}
\newcommand{\mcc}[2]{\multicolumn{#1}{c}{#2}}
\newcommand{\mcl}[2]{\multicolumn{#1}{|c}{#2}}
\newcommand{\mcr}[2]{\multicolumn{#1}{c|}{#2}}
\newcommand{\mclr}[2]{\multicolumn{#1}{|c|}{#2}}
\newcommand{\mc}{\multicolumn{1}{c}}
\newcommand{\ml}{\multicolumn{1}{|c}}
\newcommand{\mr}{\multicolumn{1}{c|}}
\newcommand{\mlr}{\multicolumn{1}{|c|}}
\newcommand{\medstrut}{\rule[-7pt]{0pt}{22pt}}
\newcommand{\tbstrut}{\rule[-6pt]{0pt}{20pt}}
\newcommand{\bstrut}{\rule[-6pt]{0pt}{16pt}}
\newcommand{\tstrut}{\rule{0pt}{14pt}}
\newcommand{\pim}{\mbox{$\pi^-$}}
\newcommand{\pimi}{\mbox{$\pi^-$}}
\newcommand{\pip}{\mbox{$\pi^+$}}
\newcommand{\pipl}{\mbox{$\pi^+$}}
\newcommand{\piz}{\mbox{$\pi^0$}}
     \def\etal{{\em et al.}}

\preprint{\tighten\vbox{\hbox{\hfil CLEO CONF 00-4}
                        \hbox{\hfil ICHEP00 999}
}}
\title{Observation of the \omegac\ Charmed Baryon at CLEO} 
\author{CLEO Collaboration}
\date{\today}
\maketitle
\tighten

\begin{abstract}

The CLEO detector at the CESR collider has used 13.7 $fb^{-1}$
to search for the production of 
\omegac\ (css-ground state) in $e^{+}e^{-}$ collisions at $\sqrt{s} \simeq 10.6$ \gev.
The modes used to study the \omegac\  are \omgmi\pip, \omgmi\pip\piz, 
\casmi\kmi\pip\pip, \casz\kmi\pip, and \omgmi\pip\pip\pim.  
We observe $40.4\pm9.0(stat)$ combined events at a mass of $2694.6\pm2.6(stat)\pm2.4(syst)$
\mevcsq. We have also measured the $\sigma \cdot {\cal B}r$ of the above
modes for scaled momentum \xp\ $>$ 0.5 to be $11.3\pm3.9\pm2.3$ $fb$, $47.6\pm18.0\pm2.8$ $fb$, 
$45.1\pm23.2\pm4.1$ $fb$, $18.2\pm10.6\pm3.8$ $fb$, and
$<5.1$ $fb$ $@$ 90 $\%$ CL, respectively. The results described here are
all preliminary.
\end{abstract}
\vfill
\begin{flushleft}
.\dotfill .
\end{flushleft}
\begin{center}
Submitted to XXXth International Conference on High Energy Physics, July
2000, Osaka, Japan
\end{center}

\newpage
{
\renewcommand{\thefootnote}{\fnsymbol{footnote}}

\begin{center}
S.~Ahmed,$^{1}$ M.~S.~Alam,$^{1}$ S.~B.~Athar,$^{1}$
L.~Jian,$^{1}$ L.~Ling,$^{1}$ M.~Saleem,$^{1}$ S.~Timm,$^{1}$
F.~Wappler,$^{1}$
A.~Anastassov,$^{2}$ J.~E.~Duboscq,$^{2}$
% ...
\end{center}
 
\small
\begin{center}
$^{1}${State University of New York at Albany, Albany, New York 12222}\\
$^{2}${Ohio State University, Columbus, Ohio 43210}\\
% ...
\end{center}

\setcounter{footnote}{0}
}
\newpage

 Various experimental groups have published results for \omegac\ in many decay 
modes, but the results are ambiguous. The WA62 experiment~\cite{wa62}, 
claimed the first evidence of \omegac\ in the \casmi\kmi\pip\pip\ decay mode with a mass of
$2746.0\pm20.0$~\mevcsq. The ARGUS Collaboration\cite{arg}, published a \omegac\ 
signal in the \casmi\kmi\pip\pip\ mode, with a mass of $2719.0\pm7.0\pm2.5~\mevcsq$ 
and $\sigma\cdot {\cal B}r$ of $2.41\pm0.90\pm0.30$ $pb$ using $0.380$ $fb^{-1}$ of
integrated luminosity. The result was contradicted by CLEO (using $1.8fb^{-1}$)
in an unpublished conference paper~\cite{conf}. Later, E687\cite{e687}
published \omegac\ with a mass of $2705.9\pm3.3\pm2.3$~\mevcsq\ in the \omgmi\pip\ mode
and a significant signal at $2699.9\pm1.5\pm2.5$~\mevcsq\ in the \sigpl\kmi\kmi\pipl\ decay mode. 
In 1995, the WA89 Collaboration\cite{wa89} reported 200 \omegac\ events in seven decay modes,
with an average mass of $2707.0\pm1.0(stat)$~\mevcsq; WA89 never published the 
\omegac\ mass.
 
 The \omegac\ (\cq\{\sq\sq\}) is a 
$J^{p}$ = $\frac{1}{2}^{+}$ ground state baryon, where \{\sq\sq\} denotes the symmetric
nature of its wave function with respect to the interchange of light-quark spins.
Different theoretical models\cite{RLP,jen,sam,am} predict the \omegac\ mass in a range from 
2664 - 2786~\mevcsq.

 The data used in this analysis were collected with CLEO II\cite{yk} and the upgraded
CLEO II.V\cite{hill} detector operating at the Cornell Electron Storage Ring (CESR), and
correspond to an integrated luminosity of 13.7 $fb^{-1}$ from the $\Upsilon(4S)$
resonance and the continuum region at energies just below. We searched for the \omegac\ 
in the five decay modes \omgmi\pip, \omgmi\pip\piz, 
\omgmi\pip\pip\pim, \casmi\kmi\pip\pip, and \casz\kmi\pip. These five
modes were chosen as most likely to show an \omegac\ signal, based upon
the pattern of other charmed baryon decays, considerations of detector
efficiency, and the size of the combinatorial backgrounds.
A sixth channel, \sigpl\kmi\kmi\pip,
was also investigated because E687\cite{e687} showed a 
significant signal in this decay mode.

 Charmed baryons at CESR are either produced from the secondary decays of $B$ mesons
or directly from e$^{+}$e$^{-}$ annihilations to \cq\cbar\ jets. We introduce \xp\ as
the scaled momentum of a \omegac\ candidate, where $\xp = p/p_{\rm max}$, and $p_{\rm max}$ =
$\sqrt{E^{2}_{\rm b} - m^{2}}$ with $E_{\rm b}$ equal to the beam energy and 
$m$ the mass of the \omegac\ candidate.
Our search is limited to \xp\ $>$ $0.5$ or \xp\ $>$ $0.6$, depending on decay mode, 
to avoid the combinatorial 
backgound that dominates at low \xp. Charmed baryons from $B$ meson decays
are kinematically limited to \xp\ $<$ 0.5, so our search is
limited to the \omegac\ baryons produced by $e^{+}e^{-}$ continuum. 
We implemented $p/k/\pi$ identification by means of a joint probability for 
the $p/k/\pi$ hypotheses by combining the specific ionization ($dE/dx$) in the wire
drift chamber and the time-of-flight in the scintillation counters. A charged track
is defined to be consistent with a particular particle hypothesis if the corresponding
probability is greater than $0.1\%$.

 We begin by reconstructing \lz\decays $p$\pim, \casz\decays\lz\piz,
\casmi\decays\lz\pim, \omgmi\decays\lz\kmi, and \sigpl\decays$p$\piz.
The analysis procedure for reconstructing
these particles closely follows that presented earlier, \cite{paul,jessop,jim}.
The hyperons were required to have vertices well separated from the beamspot,
with the flight distance of the \lz\ greater than that of the \casz, \casmi, 
or the \omgmi. We then combine these hyperons with tracks from the
primary event vertex to reconstruct \omegac\ candidates.
 Below we present \omegac\ reconstruction in the six decay modes
described above. 

\begin{figure}
\unitlength 1in
\begin{picture}(6.0,7.0)(0.0,0.0)
\put(-1.3,-2.1){\psfig{file=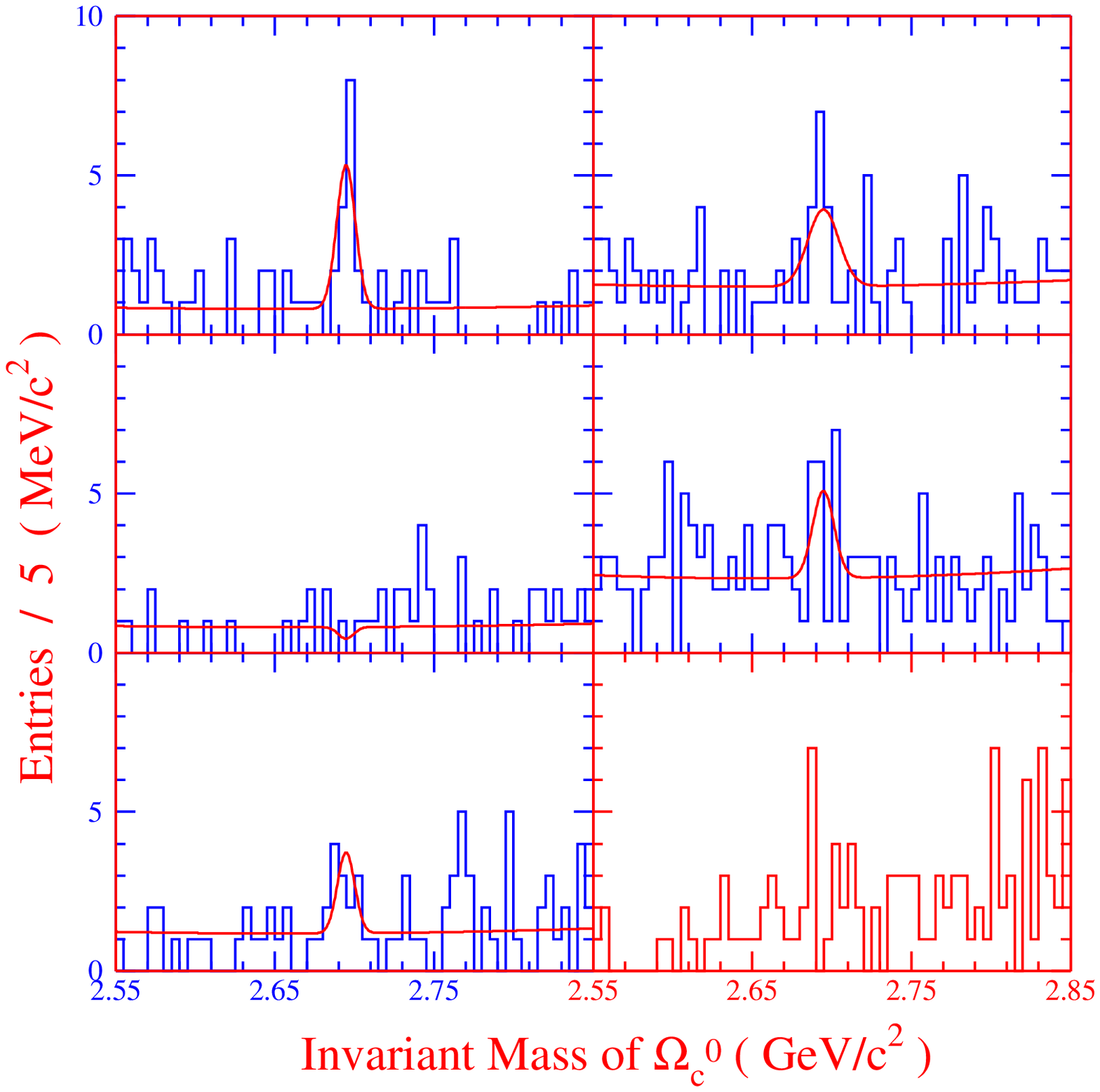,width=9.0in,bbllx=0,bblly=0,bburx=600,bbury=400}}
\put(0.6,6.50){\begin{small} (a) \end{small}}
\put(3.6,6.50){\begin{small} (b) \end{small}}
\put(0.6,4.50){\begin{small} (c) \end{small}}
\put(3.6,4.50){\begin{small} (d) \end{small}}
\put(0.6,2.40){\begin{small} (e) \end{small}}
\put(3.6,2.40){\begin{small} (f) \end{small}}
\end{picture}
\caption[]{
The above plot shows simultaneous fits to the five \omegac\ modes: (a) \omgmi\pip, 
(b) \omgmi\pip\piz, (c) \omgmi\pip\pip\pim, (d) \casz\kmi\pip, (e) \casmi\kmi\pip\pip.
The mode (f) \sigpl\kmi\kmi\pip\ has not been included in the fit.
}
\label{fig:omczo}
\end{figure}

\begin{figure}
\unitlength 1in
\begin{picture}(6.0,7.0)(0.0,0.0)
\put(0.0,0.0){\psfig{file=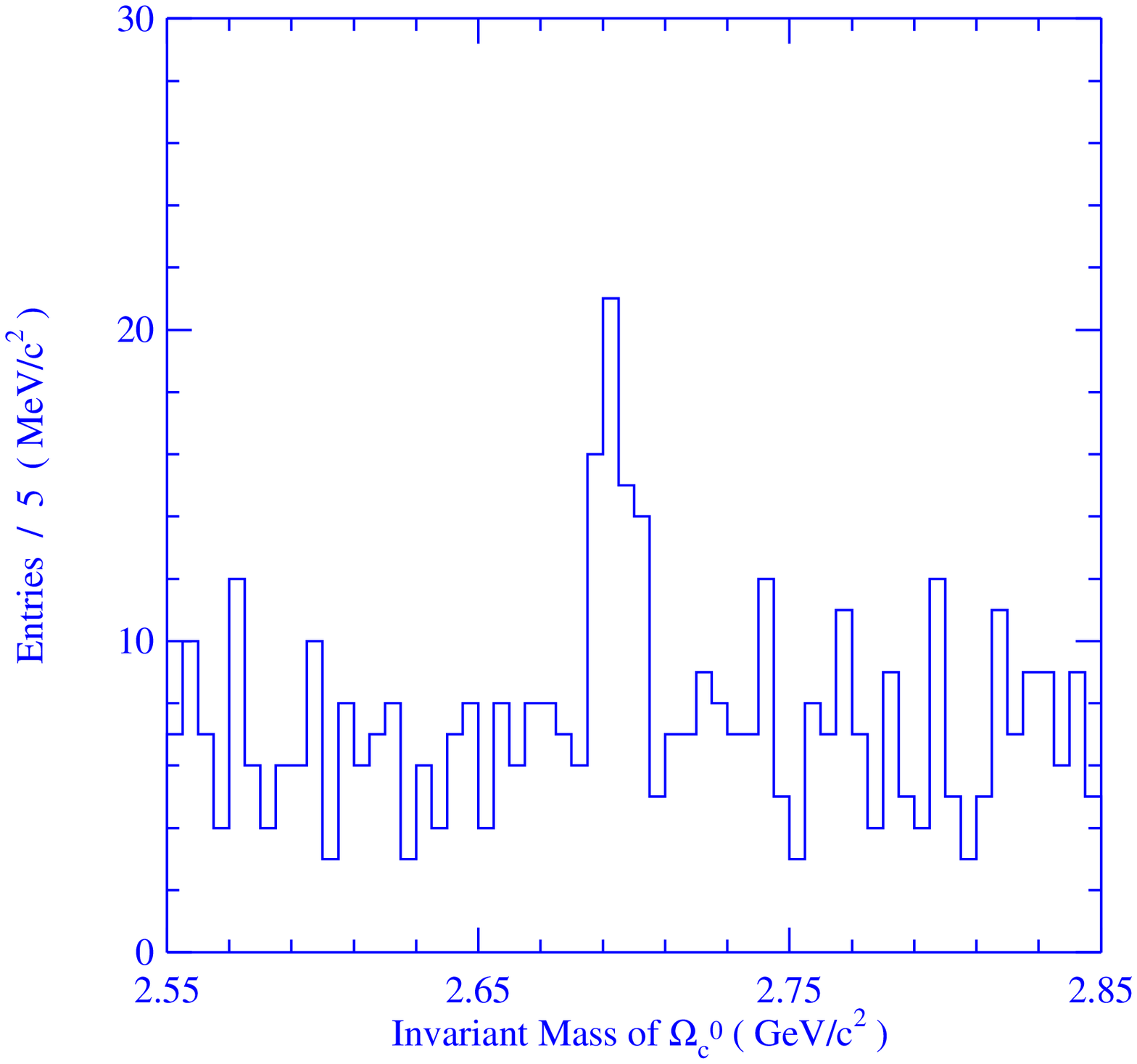,width=7.0in,bbllx=0,bblly=0,bburx=600,bbury=500}}
\end{picture}
\caption[]{
The summed plot for \omgmi\pip, \omgmi\pip\piz, \omgmi\pip\pip\pim,  
\casz\kmi\pip, and \casmi\kmi\pip\pip. 
}
\label{fig:sum}
\end{figure}

In all modes, the signal area above the background is obtained by fitting with a sum of a
Gaussian signal function (with widths fixed at signal Monte Carlo predicted values) and a second
order polynomial background. Charge conjugation is implied throughout the analysis. 
In the \omgmi\pip\ mode, we required \xp\ to be greater
than 0.5 and the \pip\ momentum to be greater than 0.5 \gevc. Figure~\ref{fig:omczo}(a)
shows the invariant mass distribution; a fit to this distribution yields a
signal of $13.3\pm4.1$ events. In the \omgmi\pip\piz\ mode, we assume the
photons used for reconstructing \piz\decays\phot\phot\ come from the
event vertex. Only \phot\phot\ combinations having invariant mass within
12.5 \mevcsq\ ($2.5\sigma$) of the nominal mass are used as \piz\ candidates.
Figure~\ref{fig:omczo}(b) shows the invariant mass distribution. Here we 
required \xp\ to be greater than 0.5 and the \pip\ and \piz\ momenta to be greater than
0.3 and 0.5 \mevc, respectively. The fit gives a yield of $11.8\pm4.9$ events.
Figure~\ref{fig:omczo}(c) shows the \omgmi\pip\pim\pip\ invariant mass distribution
for \xp\ greater than 0.5. All the charged pions are required
to have momenta greater than 0.2 \mevc. The fit yields a signal of $-0.9\pm1.4$ events.
In the \casz\kmi\pip\ mode, we considered combinations with \xp\ greater than 0.6,
since combinatorial background is higher in this mode. Figure~\ref{fig:omczo}(d)
shows the invariant mass distribution with a fit 
yielding a signal of $9.2\pm4.9$ events. In the \casmi\kmi\pip\pip\ 
mode, we required \xp\ to be greater than 0.6 and pion and kaon momenta 
to be greater than 0.2 and 0.3 \gevc, respectively. A fit to the \casmi\kmi\pip\pip\ 
distribution yields a signal of  $7.0\pm3.7$ events.
Finally, in the \sigpl\kmi\kmi\pip\ mode, we required \xp\ to be greater
than 0.5 and required charged track momenta to be greater than 0.3 \gevc. We
find the yield to be $<$ 9.5 $@$ 90 $\%$ C.L. Figure~\ref{fig:omczo}(f)
shows the invariant mass distribution for \sigpl\kmi\kmi\pip\ mode. 
The efficiency for \sigpl\kmi\kmi\pip\ reconstruction is $\sim 15\%$ of
that for the \omgmi\pip\ mode, our highest-yield. We have not included the \sigpl\kmi\kmi\pip\ 
mode in the mass measurement.
The total yield in five modes combined, excluding \sigpl\kmi\kmi\pip, sums to $40.4\pm9.0$, 
as shown in Table~\ref{tab:yield}. The mass distribution for the
five modes combined is shown in Figure~\ref{fig:sum}.

 To determine the mass, we have performed an unbinned maximum-likelihood 
fit using the sum of a single 
Gaussian and a second order polynomial background. There are
two inputs to the fit, the invariant mass $M_i$ and the corresponding
mass resolution $\sigma_i$ of each mass candidate from 2.55 to 2.85 \gevcsq. 
The likelihood function to maximize is the product of probability density
functions (PDFs) for all the
candidate events, and has the following form:
\begin{equation}
  {\cal L}(M(\omegac),f_s,a_1,a_2) = \prod_{i} \left[ f_s G(M_i - M(\omegac)|S\sigma_i)
   + (1 - f_s)\frac{P(M_i)}{\int^{2.85}_{2.55}P(M_i)dM_i} \right],
\end{equation}
 where G($y | \sigma$) = (1/$\sqrt{2\pi\sigma}$)exp($-y^2/2\sigma^2$) 
and P($y$) = $1.0 + a_1(y-2.7) + a_2(y-2.7)^2$. $M(\omegac)$ is the fitted 
\omegac\ mass, $S$ is the global scale factor multiplying $\sigma_i$, and
$f_s$ is the fraction of signal events under G($y | \sigma$). The fitted
mass for the above PDF is $2694.9\pm0.1$ \mevcsq\ for the
Monte Carlo and $2694.6\pm2.6$ \mevcsq\ for the data. The \omegac\ Monte Carlo was
generated at a mass of 2695 \mevcsq.
The fitted scale factor $S$ is $1.72\pm 0.42$ for the data
and $1.16\pm0.02$ for the simulated events.

 We have also checked for goodness-of-fit by performing 
ten different ``toy'' Monte Carlo experiments. In each experiment 
we took sideband events from the wrong sign combinations in the data and 
signal events from the Monte Carlo. The $-2\ln{\cal L}$ of the fit ranged from 518
to 576; the $-2\ln{\cal L}$ of the fit to the data is 564.
Twenty percent of the experiments have greater $-2\ln{\cal L}$ than the data. 

 We also studied the momentum spectrum of \omegac, finding consistency
with that for other charmed baryons~\cite{edw}.

The mass calibration of our detector was checked by the \cascz, 
which has similar spectator decay modes with the same number of charged tracks
in the final state as the \omegac.
The mass of the reconstructed \cascz\ from the \piz\ mode is
lower than from the all-charged modes. The asymmetric \piz\ mass peak,
due to the mismeasured photons at low energies, accounts for this 
low mass. The mass difference for
\cascz\ with and without \piz\ involved in the final state is 2.0 \mevcsq.
The \lcpl\ mass, studied in different decay modes, shows a spread of 1.3 \mevcsq. 
Adding these in quadrature,
we assign a total systematic error of 2.4 \mevcsq\ to our \omegac\ mass measurement. 

 We have also measured $\sigma \cdot {\cal B}r$ for \omgmi\pip, \omgmi\pip\piz, 
\casmi\kmi\pip\pip, \casz\kmi\pip,\omgmi\pip\pip\pim and \sigpl\kmi\kmi\pip\
to be $11.3\pm3.9\pm2.3$ $fb$, $47.6\pm18.0\pm2.8$ $fb$, 
$45.1\pm23.2\pm4.1$ $fb$, $18.2\pm10.6\pm3.8$ $fb$, $<5.1$ $fb$ $@$ 90 $\%$ CL, and
$< 53.8$ $fb$ $@$ 90 $\%$~C.L. $fb$, respectively, as shown in Table~\ref{tab:yield}.
We estimated the systematic errors for the branching fraction by changing the
\omegac\ mass by $\pm 1.0 \sigma$ from its best fit value.

\begin{table}[h]
\begin{center}
\begin{small}
\caption{ Fit to mass distribution. }
\label{tab:yield}
\begin{tabular}{c c c c c} \hline
\multicolumn{1}{c}{ }&$\sigma_{MC}(\mevcsq)$& Fitted Yield& Relative ${\cal B}r$& $\sigma \cdot {\cal B}r$ (fb) \\ 
                     &               &mode dependent \xp\ &all \xp\ $>$ 0.5&all \xp\ $>$ 0.5\\ \hline
\omgmi\pip\          & 5.87                 &13.3$\pm$4.1 &1.0             &$11.3\pm3.9\pm2.3$ \\
\omgmi\pip\piz       & 9.71                 &11.8$\pm$4.9 &$4.2\pm2.2\pm1.2$&$47.6\pm18.0\pm2.8$\\
\casz\pip\kmi        & 6.72                 &9.2$\pm$4.9  &$4.0\pm2.5\pm0.6$&$45.1\pm23.2\pm4.1$ \\
\casmi\pip\pip\kmi   & 5.46                 &7.0$\pm$3.7  &$1.6\pm1.1\pm0.4$&$18.2\pm10.6\pm3.8$\\ 
\omgmi\pip\pip\pimi  & 4.89                 &-0.9$\pm$1.4 & $<$ 0.56    & $<$ 5.1  $@$ 90 $\%$ CL  \\
Combined 5 modes     &                      &40.4$\pm$9.0 &              &  \\ 
\sigpl\kmi\kmi\pip   & 6.18                 &2.8$\pm$4.1  & $<$ 4.8      & $<$ 53.8 $@$ 90 $\%$ CL \\ \hline
\end{tabular}
\end{small}
\end{center}
\end{table}               

 In conclusion, we observe a narrow resonance with a mass around $2694.6\pm2.6 
\pm2.4$ \mevcsq\ in five decay modes \omgmi\pip, \omgmi\pip\piz, 
\omgmi\pip\pip\pim, \casmi\kmi\pip\pip, and \casz\kmi\pip. Although 
the signal is not statistically significant in any individual mode, the
combined signal stands out over the background with a yield of $40.4\pm9.0(stat)$
events.

\section*{ACKNOWLEDGMENTS} 
We gratefully acknowledge the effort of the CESR staff in providing us with
excellent luminosity and running conditions.
I.P.J. Shipsey thanks the NYI program of the NSF, 
M. Selen thanks the PFF program of the NSF, 
A.H. Mahmood thanks the Texas Advanced Research Program,
M. Selen and H. Yamamoto thank the OJI program of DOE, 
M. Selen and V. Sharma 
thank the A.P. Sloan Foundation, 
M. Selen and V. Sharma thank the Research Corporation, 
F. Blanc thanks the Swiss National Science Foundation, 
and H. Schwarthoff and E. von Toerne
thank the Alexander von Humboldt Stiftung for support.  
This work was supported by the National Science Foundation, the
U.S. Department of Energy, and the Natural Sciences and Engineering Research 
Council of Canada.

\pagebreak

\clearpage
\pagebreak
\clearpage
\pagebreak
\end{document}